\documentclass{aa}
\usepackage[dvips]{graphicx}  
\usepackage{txfonts}
\usepackage{lscape}
\usepackage{color} 
\begin{document}
\newcommand{\Vsini}{$V \sin i\ $}
\newcommand{\MJup}{$M_{\textrm{\footnotesize Jup}}$}
\newcommand{\Msini}{$M \sin i$}
\newcommand{\Msun}{$M_\odot$}
\newcommand{\Rsun}{$R_\odot$}
\newcommand{\Lsun}{$L_\odot$}
\newcommand{\teff}{$T_{\textrm{eff}}$}
\newcommand{\feh}{$\left[\right.$Fe/H$\left.\right]$}
        \title{On the chromospheric activity of stars with planets}
        \author{B. L. Canto Martins
                                \and M. L. das Chagas
                                \and S. Alves
                                \and I. C. Le\~ao
                                \and L. P. de Souza Neto
                                \and J. R. de Medeiros
                            }
\offprints{B. L. Canto Martins \\ \email{brunocanto@dfte.ufrn.br}}
        \institute{ Departamento de F\'isica,
Universidade Federal do Rio Grande do
Norte, Natal, Brazil\\ 
                                }
        \date{Received ----------; accepted ----------}
        \abstract
        {Signatures of chromospheric activity enhancement have been found for a dozen stars, 
pointing to a possible star-planet interaction. Nevertheless in the coronal activity regime, there is no 
conclusive observational evidence for such an interaction. Does star-planet interaction manifest itself only for 
a few particular cases, without having a major effect on stars with planets in general? }
        {We aim to add additional observational constraints to support or reject the major effects of star-planet interactions 
in stellar activity, based on CaII chromospheric emission flux. }
        {We performed a statistical analysis of CaII emission flux of stars with planets, as well as a comparison 
between CaII  and X-ray emission fluxes, searching for dependencies on planetary parameters.}
        {In the present sample of stars with planets, there are no significant correlations between chromospheric activity indicator $\log (R'_{HK})$
and planetary parameters. Further, the distribution of the chromospheric activity indicator for stars without planets is not 
distinguishable from the one with planets.}
        {}
\keywords{Planet--star interaction -- Stars: activity -- Stars: Chromosphere -- Stars: coronae -- Stars: statistics}
\maketitle
\authorrunning{B. L. Canto Martins:}
\titlerunning{On the chromospheric activity of stars with planets}
\hyphenation{e-vo-lu-ti-o-na-ry ac-ti-vi-ty res-pon-si-ble ve-lo-ci-ty
lu-mi-no-si-ty lea-ding}

\section{Introduction}

Since the pioneering discovery by Michel Mayor and Didier
Queloz (1995), a little more than a decade ago, of a planet orbiting the star {\it 51
Peg}, more than 450 other exoplanets have been found at the time of
writing (e.g.: Schneider 2010). The discovered planets have masses ranging from 4 Earth
masses to 11 Jupiter masses. They can be found at distances of several AU or
close to the parent star, with orbital periods ranging from a few days to a
few years. High eccentricity is a common parameter connecting these
planets (e.g.: Marcy et al. 2001). 

These discoveries have inspired intensive studies on the physical properties of the 
planets and of their parent stars, including a possible star-planet interaction.
Indeed, in an analogy to binary stars, which show a higher activity level compared to single stars,
at very close distances, one might also expect planets to play a role in the level
of activity of their host stars. Nevertheless, observational data, yet at a very informative level, demonstrate
that star-planet interaction appears to be more complex than the widely accepted star-star interaction in close binary systems. 
For instance, Shkolnik et al. (2003, 2008) reported a planet-induced chromospheric activity 
on two stars with planets, HD 179949 and $\upsilon~And$, apparent from the night-to-night modulation of the CaII H and K 
chromospheric emission phased with the hot Jupiter's orbit.  In addition, Kashyap et al. (2008) claimed that stars with 
close-in giant planets are on average more X-ray active than those with planets that are more distant, an observational result 
consistent with the hypothesis that giant planets in close proximity to the parent stars could influence stellar magnetic activity.

By contrast, Poppenhaeger et al. (2010) found no significant correlations between X-ray luminosity and planetary parameters, 
suggesting no major average activity enhancement in the corona of stars with planets. Indeed, these authors found no 
additional detectable effects in coronal X-ray luminosity that could be associated to coronal manifestations  of star-planet 
interaction. According to Poppenhaeger et al. (2010), any trends in the X-ray luminosity versus planet orbital parameters 
seem to be dominated by selection effects. In addition, Fares et al. (2010) reported no clear evidence of  star-planet magnetospheric interactions
in HD 189733. In this context, let us recall that more recently, from a comparison of the distribution of the rotation of stars with and without detected  planets, Alves et al. (2010)  showed that the $v \sin i$ distribution for these two families 
of stars is drawn from the same population distribution function.

Because the observational basis of stellar activity enhancement due to star-planet interaction is, clearly, not yet established, 
in the present study we report a statistical analysis of stellar CaII chromospheric emission flux for a 
sample of 74 stars with planets, as described in the following sections. For a more solid analysis of the behavior of chromospheric and 
coronal activity of stars with planets, it is essential to conduct a comparative analysis of stars without  detected planets. 
This is one of the major goals of the current study. In this context, we also analyzed the CaII chromospheric emission flux of 
 a comparison sample of 26 stars without detected planets. The paper is organized as follows: in Section 2 we present the characteristics 
of the working samples, in Section 3 we describe our findings, with a brief discussion and finally, we outline our conclusions in Section 4.

\section{Stellar working sample and data}

\begin{figure}
\centering
\includegraphics[width=9cm]{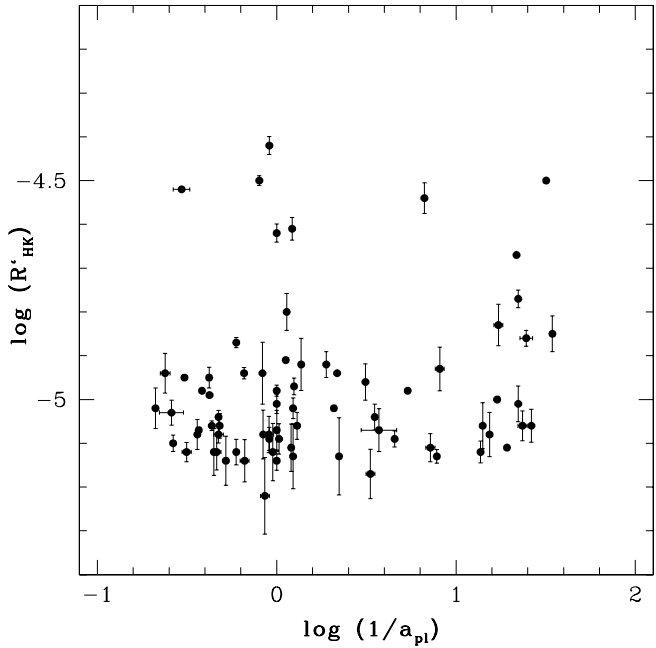}
\caption{Chromospheric activity indicator $\log (R'_{HK})$ as a function of the planetary semi-major axis $\log (1/a_{pl})$ for our sample of stars with planets.}
\label{default}
\end{figure}

Different methods, such as radial velocity measurements (Butler et al. 1996; de Medeiros et al. 2009), and photometric light curves (Konacki
et al. 2003, 2004), have been used to discovery the existence of new planetary systems around other stars than the Sun. Aiming to make our sample homogeneous, we use only stars that had their planets discovered using radial velocity method.

The present stellar working sample consists of selected objects from the base of extrasolar planets maintained by Jean Schneider (Schneider 2010), updated on October 6, 2009, when there were 235 planets cataloged by the method of variation of the radial velocity (RV). We checked which of these stars
presented Ca II emission flux detected in the catalog of Wright et al. (2004). After this selection, we filtered the complete stellar sample to
obtain a subset of main-sequence stars that are within 20 pc in the solar neighborhood. Our final working sample consists of 74 stars with planets with spectral types in the F-G-K interval, as presented in Table~1.

The chromospheric activity indicator, $\log (R'_{HK})$, for all stars was computed from the Ca II H and K line-core emission index listed by
Wright et al. (2004), following the procedure of converting emission index {\it S} to the flux at the stellar surface proposed by Noyes et al.
(1984). To diagnose coronal activity of the stars forming the above mentioned stellar sample, we used the coronal activity indicator, $\log (\frac{L_X}{L_{bol}})$, that was computed using the X-ray fluxes listed by Kashyap et al. (2008) and Poppenhaeger et al. (2010). Readers are referred to these
works for a discussion of the observational procedures, data reduction and error analysis.

Stellar luminosities were determined as follows: first, apparent visual magnitudes (m$_v$) and trigonometric parallaxes, both taken from the HIPPARCOS catalogue (ESA 1997), were combined to yield the absolute visual magnitude (M$_v$). Bolometric correction ($BC$), computed from Flower (1996) calibration, was applied to obtain the bolometric magnitude, which was finally converted into stellar luminosity. The effective temperature was computed using Flower (1996) ($B-V$) versus T$_{eff}$ calibration. The stellar and planetary parameters, as well as chromospheric and coronal activity indicators for the entire sample of stars with planets, are listed in Table 1. 


For comparative purposes, we took a sample of stars not known to have any planetary-mass companions. This new sample was taken from Schmitt et al. (1997), as done by Poppenhaeger et al. (2010), and is composed of 26 F-G-K type dwarf stars with all stars having Ca II emission flux measured by Wright et al. (2004) and X-ray fluxes detected by Schmitt (1997). The chromospheric and coronal activity indicators were calculated as before for the sample of stars with planets. However, we should be cautious about this sample, which derives from a list of stars that are surveyed for planets, but for which none have yet been found. This certainly does not mean that such stars have no planetary companions whatsoever. In fact, they might host planets with very low mass and/or a long orbital period that are more difficult to detect with radial velocity surveys. Stellar properties, as well as chromospheric and coronal activity indicators for stars without detected planets are given in Table 2.


\begin{table*}															
\caption{Stellar and planetary parameters and chromospheric and coronal activity indicators of stars with planets ({\bf Sources:} {\it a} - Kashyap et al. (2008); {\it b} Poppenhaeger et al. (2010)).}
\begin{center}															
\begin{tabular}{llcccccc} 															
\hline \hline															
	Star	&	ST	&	$(B-V)$		&	$a_{pl}$	&	$M_{pl}$	&	$\log (R'_{HK})$& $\log (\frac{L_X}{L_{bol}})^a$&	$\log (\frac{L_X}{L_{bol}})^b$\\
\hline															
HD 3651		&	K0V	&	0.850$\,\pm\,$0.009	&	0.284			&	0.200			&	-5.04$\,\pm\,$0.03	&	-6.07$\,\pm\,$0.08	&	-6.07$\,\pm\,$0.23	\\
HD 4203		&	G5	&	0.771$\,\pm\,$0.021	&	1.164$\,\pm\,$0.07	&	2.070$\,\pm\,$0.18	&	-5.22$\,\pm\,$0.09	&	-5.18$\,\pm\,$0.22	&				\\
HD 4208		&	G5V	&	0.664$\,\pm\,$0.004	&	0.800			&	1.700			&	-4.97$\,\pm\,$0.02	&	-5.12			&				\\
HD 7924		&	K0	&	0.826$\,\pm\,$0.006	&	0.029			&	0.057			&	-4.85$\,\pm\,$0.04	&				&	-5.71$\,\pm\,$0.29	\\
HD 8574		&	F8	&	0.577$\,\pm\,$0.011	&	0.770			&	2.110			&	-5.06$\,\pm\,$0.03	&	-5.00			&				\\
HD 10697	&	G5IV	&	0.720$\,\pm\,$0.009	&	2.160$\,\pm\,$0.12	&	6.380$\,\pm\,$0.53	&	-5.12$\,\pm\,$0.04	&	-5.83$\,\pm\,$0.17	&				\\
HD 12661	&	K0	&	0.710$\,\pm\,$0.015	&	0.830			&	2.300			&	-5.11$\,\pm\,$0.05	&	-5.44			&				\\
HD 16141	&	G5IV	&	0.670$\,\pm\,$0.004	&	0.350			&	0.215$\,\pm\,$0.03	&	-5.14$\,\pm\,$0.02	&	-5.32			&				\\
HD 19994	&	F8V	&	0.575$\,\pm\,$0.005	&	1.680			&	1.420			&	-4.87$\,\pm\,$0.01	&	-6.04			&	-6.01$\,\pm\,$0.28	\\
HD 20367	&	G0	&	0.574$\,\pm\,$0.008	&	1.250			&	1.070			&	-4.50$\,\pm\,$0.01	&	-4.54$\,\pm\,$0.03	&	-4.42$\,\pm\,$0.12	\\
HD 23596	&	F8	&	0.634$\,\pm\,$0.009	&	2.720			&	7.190			&	-5.07			&				&	-5.34			\\
HD 30562	&	F8V	&	0.631$\,\pm\,$0.003	&	2.300$\,\pm\,$0.02	&	1.290$\,\pm\,$0.08	&	-5.06$\,\pm\,$0.01	&				&	-7.08			\\
HD 37124	&	G4IV-V	&	0.667$\,\pm\,$0.008	&	0.529$\,\pm\,$0.00	&	0.640$\,\pm\,$0.05	&	-4.92$\,\pm\,$0.03	&	-5.15			&				\\
HD 40979	&	F8	&	0.573$\,\pm\,$0.007	&	0.811			&	3.280$\,\pm\,$3.00	&	-4.62$\,\pm\,$0.02	&	-5.51			&				\\
HD 45350	&	G5	&	0.740$\,\pm\,$0.015	&	1.920$\,\pm\,$0.07	&	1.790$\,\pm\,$0.14	&	-5.14$\,\pm\,$0.06	&	-5.11			&				\\
HD 46375	&	K1IV	&	0.860$\,\pm\,$0.000	&	0.041			&	0.249$\,\pm\,$0.03	&	-4.98$\,\pm\,$0.01	&	-6.02$\,\pm\,$0.08	&				\\
HD 49674	&	G0	&	0.729$\,\pm\,$0.015	&	0.058$\,\pm\,$0.00	&	0.115$\,\pm\,$0.02	&	-4.83$\,\pm\,$0.05	&	-5.02			&				\\
HD 50499	&	G1V	&	0.614$\,\pm\,$0.008	&	3.860$\,\pm\,$0.60	&	1.710$\,\pm\,$0.20	&	-5.03$\,\pm\,$0.03	&	-4.98			&				\\
HD 50554	&	F8	&	0.582$\,\pm\,$0.008	&	2.380			&	4.900			&	-4.95$\,\pm\,$0.02	&	-6.81			&				\\
HD 52265	&	G0III-IV&	0.572$\,\pm\,$0.003	&	0.490			&	1.050$\,\pm\,$0.03	&	-5.01$\,\pm\,$0.02	&	-5.60			&	-6.65$\,\pm\,$0.17	\\
HD 66428	&	G5	&	0.715$\,\pm\,$0.002	&	3.180$\,\pm\,$0.19	&	2.820$\,\pm\,$0.03	&	-5.12$\,\pm\,$0.02	&	-4.87			&				\\
HD 68988	&	G0	&	0.652$\,\pm\,$0.015	&	0.071			&	1.900			&	-5.06$\,\pm\,$0.05	&	-4.93			&				\\
HD 69830	&	K0V	&	0.754$\,\pm\,$0.009	&	0.186			&	0.038			&	-4.98			&	-5.89$\,\pm\,$0.15	&	-5.89$\,\pm\,$0.30	\\
HD 72659	&	G0	&	0.612$\,\pm\,$0.015	&	4.740$\,\pm\,$0.08	&	3.150$\,\pm\,$0.01	&	-5.02$\,\pm\,$0.05	&	-5.16			&				\\
HD 74156	&	G0	&	0.585$\,\pm\,$0.014	&	0.294			&	1.880			&	-5.07$\,\pm\,$0.05	&	-5.26			&				\\
HD 80606	&	G5	&	0.765$\,\pm\,$0.025	&	0.449$\,\pm\,$0.01	&	3.940$\,\pm\,$0.11	&	-5.13$\,\pm\,$0.09	&	-4.82			&				\\
HD 82943	&	G0	&	0.623$\,\pm\,$0.001	&	0.730			&	0.880			&	-4.92$\,\pm\,$0.06	&	-5.76			&	-5.76			\\
HD 83443	&	K0V	&	0.811$\,\pm\,$0.003	&	0.041$\,\pm\,$0.00	&	0.400$\,\pm\,$0.03	&	-4.86$\,\pm\,$0.02	&	-4.96			&				\\
HD 89307	&	G0V	&	0.594$\,\pm\,$0.003	&	3.270$\,\pm\,$0.07	&	1.780$\,\pm\,$0.13	&	-4.95			&	-5.60			&				\\
HD 89744	&	F7V	&	0.531$\,\pm\,$0.003	&	0.890			&	7.990			&	-4.91			&	-6.31$\,\pm\,$0.07	&				\\
HD 92788	&	G5	&	0.694$\,\pm\,$0.005	&	0.970			&	3.860			&	-5.09$\,\pm\,$0.03	&	-5.53			&				\\
HD 99109	&	K0	&	0.874$\,\pm\,$0.002	&	1.105$\,\pm\,$0.07	&	0.502$\,\pm\,$0.07	&	-5.08$\,\pm\,$0.04	&	-5.13			&				\\
HD 99492	&	K2V	&	1.002$\,\pm\,$0.012	&	0.123$\,\pm\,$0.01	&	0.109$\,\pm\,$0.01	&	-4.93$\,\pm\,$0.05	&	-5.55$\,\pm\,$0.18	&	-6.13$\,\pm\,$0.16	\\
HD 106252	&	G0	&	0.635$\,\pm\,$0.007	&	2.610			&	6.810			&	-4.98			&	-6.03			&				\\
HD 107148	&	G5	&	0.707$\,\pm\,$0.013	&	0.269$\,\pm\,$0.06	&	0.210$\,\pm\,$0.04	&	-5.07$\,\pm\,$0.05	&	-5.35			&				\\
HD 108874	&	G5	&	0.738$\,\pm\,$0.018	&	1.051$\,\pm\,$0.02	&	1.360$\,\pm\,$0.13	&	-5.12$\,\pm\,$0.07	&	-5.14			&				\\
HD 114729	&	G0V	&	0.591$\,\pm\,$0.008	&	2.080			&	0.820			&	-5.06$\,\pm\,$0.03	&	-5.66			&				\\
HD 114783	&	K0	&	0.930$\,\pm\,$0.013	&	1.200			&	0.990			&	-4.94$\,\pm\,$0.07	&	-5.49			&	-6.60$\,\pm\,$0.19	\\
HD 117207	&	G8IV/V	&	0.724$\,\pm\,$0.002	&	3.780			&	2.060			&	-5.10$\,\pm\,$0.02	&	-5.47			&				\\
HD 128311	&	K0	&	0.973$\,\pm\,$0.004	&	1.099$\,\pm\,$0.04	&	2.180$\,\pm\,$0.02	&	-4.42$\,\pm\,$0.02	&	-4.54$\,\pm\,$0.06	&	-4.54$\,\pm\,$0.21	\\
HD 130322	&	K0III	&	0.781$\,\pm\,$0.002	&	0.880			&	1.080			&	-4.80$\,\pm\,$0.04	&	-5.69$\,\pm\,$0.07	&	-5.66$\,\pm\,$0.16	\\
HD 134987	&	G5V	&	0.691$\,\pm\,$0.020	&	0.810$\,\pm\,$0.02	&	1.590$\,\pm\,$0.02	&	-5.13$\,\pm\,$0.07	&	-5.78			&	-5.78			\\
HD 141937	&	G2/G3V	&	0.628$\,\pm\,$0.002	&	1.520			&	9.700			&	-4.94$\,\pm\,$0.01	&	-5.39			&				\\
HD 150706	&	G0	&	0.607$\,\pm\,$0.005	&	0.820			&	1.000			&	-4.61$\,\pm\,$0.03	&	-4.68$\,\pm\,$0.04	&	-4.68$\,\pm\,$0.19	\\
HD 154345	&	G8V	&	0.728$\,\pm\,$0.005	&	4.190$\,\pm\,$0.26	&	0.947$\,\pm\,$0.09	&	-4.94$\,\pm\,$0.05	&	-5.59			&	-6.02$\,\pm\,$0.16	\\
HD 164922	&	K0V	&	0.799$\,\pm\,$0.005	&	2.110$\,\pm\,$0.13	&	0.360$\,\pm\,$0.05	&	-5.08$\,\pm\,$0.02	&	-5.49			&	-6.60			\\
HD 168443	&	G5	&	0.724$\,\pm\,$0.014	&	0.300$\,\pm\,$0.02	&	8.020$\,\pm\,$0.65	&	-5.17$\,\pm\,$0.06	&	-5.50			&				\\
HD 168746	&	G5	&	0.713$\,\pm\,$0.015	&	0.065			&	0.230			&	-5.08$\,\pm\,$0.05	&	-4.96			&				\\
HD 169830	&	F8V	&	0.517$\,\pm\,$0.004	&	0.810			&	2.880			&	-5.02$\,\pm\,$0.02	&	-6.10$\,\pm\,$0.04	&				\\
HD 170469	&	G5	&	0.677$\,\pm\,$0.014	&	2.240			&	0.670			&	-5.12$\,\pm\,$0.05	&	-3.01			&				\\
HD 178911B	&	G1V	&	0.643$\,\pm\,$0.007	&	0.320			&	6.292$\,\pm\,$0.06	&	-4.96$\,\pm\,$0.04	&	-5.56$\,\pm\,$0.18	&				\\
HD 179949	&	F8V	&	0.548$\,\pm\,$0.009	&	0.045$\,\pm\,$0.00	&	0.950$\,\pm\,$0.04	&	-4.77$\,\pm\,$0.02	&	-5.25$\,\pm\,$0.10	&	-5.25$\,\pm\,$0.25	\\
HD 183263	&	G2IV	&	0.678$\,\pm\,$0.012	&	1.510$\,\pm\,$0.09	&	3.670$\,\pm\,$0.30	&	-5.14$\,\pm\,$0.05	&	-4.72			&				\\
HD 187123	&	G5	&	0.661$\,\pm\,$0.010	&	0.043$\,\pm\,$0.00	&	0.520$\,\pm\,$0.04	&	-5.06$\,\pm\,$0.03	&	-6.40			&				\\
HD 188015	&	G5IV	&	0.727$\,\pm\,$0.010	&	1.190			&	1.260			&	-5.08$\,\pm\,$0.06	&	-4.78			&				\\
HD 189733	&	G5	&	0.932$\,\pm\,$0.008	&	0.031$\,\pm\,$0.00	&	1.150$\,\pm\,$0.04	&	-4.50			&	-4.74$\,\pm\,$0.14	&	-4.91$\,\pm\,$0.19	\\
HD 190360	&	G6IV+...&	0.749$\,\pm\,$0.001	&	0.128$\,\pm\,$0.00	&	0.057$\,\pm\,$0.02	&	-5.13$\,\pm\,$0.02	&	-6.71			&	-7.45$\,\pm\,$0.21	\\
HD 192263	&	K0	&	0.938$\,\pm\,$0.015	&	0.150			&	0.720			&	-4.54$\,\pm\,$0.04	&	-5.06$\,\pm\,$0.20	&	-5.06$\,\pm\,$0.35	\\
HD 195019	&	G3IV-V&	0.662$\,\pm\,$0.007	&	0.139$\,\pm\,$0.01	&	3.700$\,\pm\,$0.30	&	-5.11$\,\pm\,$0.03	&	-7.51$\,\pm\,$0.17	&	-7.42$\,\pm\,$0.17	\\
HD 196885	&	F8IV:	&	0.559$\,\pm\,$0.006	&	2.370$\,\pm\,$0.02	&	2.580$\,\pm\,$0.16	&	-4.99			&	-5.59			&				\\
HD 209458	&	F8	&	0.594$\,\pm\,$0.015	&	0.045$\,\pm\,$0.00	&	0.690$\,\pm\,$0.02	&	-5.01$\,\pm\,$0.04	&	-6.77$\,\pm\,$0.20	&				\\
HD 210277	&	G0	&	0.773$\,\pm\,$0.006	&	1.100$\,\pm\,$0.02	&	1.230$\,\pm\,$0.03	&	-5.09$\,\pm\,$0.03	&	-5.74			&	-5.74			\\
HD 216770	&	K0V	&	0.821$\,\pm\,$0.004	&	0.460			&	0.650			&	-4.94			&	-4.95			&				\\
HD 217107	&	G8IV	&	0.744$\,\pm\,$0.006	&	0.073$\,\pm\,$0.00	&	1.330$\,\pm\,$0.05	&	-5.12$\,\pm\,$0.02	&	-6.15$\,\pm\,$0.16	&	-6.81			\\
$14\,Her$		&	K0V	&	0.877$\,\pm\,$0.006	&	2.770$\,\pm\,$0.05	&	4.640$\,\pm\,$0.19	&	-5.08$\,\pm\,$0.03	&	-6.13			&	-6.34$\,\pm\,$0.14	\\
$16\,Cyg$		&	G2V	&	0.643$\,\pm\,$0.006	&	1.680$\,\pm\,$0.03	&	1.680$\,\pm\,$0.07	&	-5.12$\,\pm\,$0.03	&	-5.90			&	-7.33			\\
\hline	
\end{tabular}															
\label{fluxes}															
\end{center}															
\end{table*}															

\begin{table*}														
{\bf Table 1.} Cont.
\begin{center}															
\begin{tabular}{llcccccc} 															
\hline \hline															
	Star	&	ST	&	$(B-V)$		&	$a_{pl}$	&	$M_{pl}$	&	$\log (R'_{HK})$& $\log (\frac{L_X}{L_{bol}})^a$&	$\log (\frac{L_X}{L_{bol}})^b$\\
\hline															
$47\,UMa$	&	G0V	&	0.624$\,\pm\,$0.003	&	2.110$\,\pm\,$0.02	&	2.600$\,\pm\,$0.07	&	-5.04$\,\pm\,$0.02	&	-6.66$\,\pm\,$0.04	&	-7.39$\,\pm\,$0.21	\\
$51\,Peg$		&	G5V	&	0.666$\,\pm\,$0.007	&	0.052			&	0.468$\,\pm\,$0.01	&	-5.11			&	-6.91$\,\pm\,$0.09	&	-7.43$\,\pm\,$0.18	\\
$55\,Cnc$		&	G8V	&	0.869$\,\pm\,$0.012	&	0.038$\,\pm\,$0.00	&	0.024$\,\pm\,$0.00	&	-5.06$\,\pm\,$0.04	&	-6.37$\,\pm\,$0.11	&	-6.32$\,\pm\,$0.14	\\
$70\,Vir$		&	G5V	&	0.714$\,\pm\,$0.007	&	0.480			&	7.440			&	-5.02			&	-6.64$\,\pm\,$0.13	&	-6.64$\,\pm\,$0.28	\\
$\epsilon\,Eri$	&	K2V	&	0.881$\,\pm\,$0.007	&	3.390$\,\pm\,$0.36	&	1.550$\,\pm\,$0.24	&	-4.52			&	-4.99$\,\pm\,$0.00	&	-4.89$\,\pm\,$0.12	\\
$\rho\,CrB$	&	G2V	&	0.612$\,\pm\,$0.003	&	0.220			&	1.040			&	-5.09$\,\pm\,$0.02	&	-6.13			&	-6.13			\\
$\tau\,Boo$	&	F7V	&	0.508$\,\pm\,$0.001	&	0.046			&	3.900			&	-4.67			&	-5.29$\,\pm\,$0.02	&	-5.18$\,\pm\,$0.12	\\
$\upsilon\,And$&	F8V	&	0.536$\,\pm\,$0.007	&	0.059$\,\pm\,$0.00	&	0.690$\,\pm\,$0.03	&	-5.00			&	-6.00$\,\pm\,$0.07	&	-6.00$\,\pm\,$0.22	\\
\hline															
\end{tabular}															
\label{fluxes}															
\end{center}															
\end{table*}															

\section{Results and discussion}

The aim of this work is  to point out additional observational constraints to support or reject major effects of star-planet interaction 
in stellar activity, based on CaII chromospheric and X-ray emission fluxes. To this end, we dedicated most of our efforts to identifying 
qualitative trends between CaII and X-ray fluxes and planetary parameters. We chose $\log (R'_{HK})$ and $\log (\frac{L_X}{L_{bol}})$ as indicators of 
chromospheric and coronal activity, respectively, because they are independent of stellar radius-induced effects. Indeed, it is expected that  any 
planet-induced activity  changes should therefore be more evident in $\log (\frac{L_X}{L_{bol}})$ (e.g.: Poppenhaeger et al.  2010). In the context of any planet-induced chromospheric activity, one should also expect more evident changes in $\log (R'_{HK})$ for such an aspect.

\begin{figure}
\centering
\includegraphics[width=9cm]{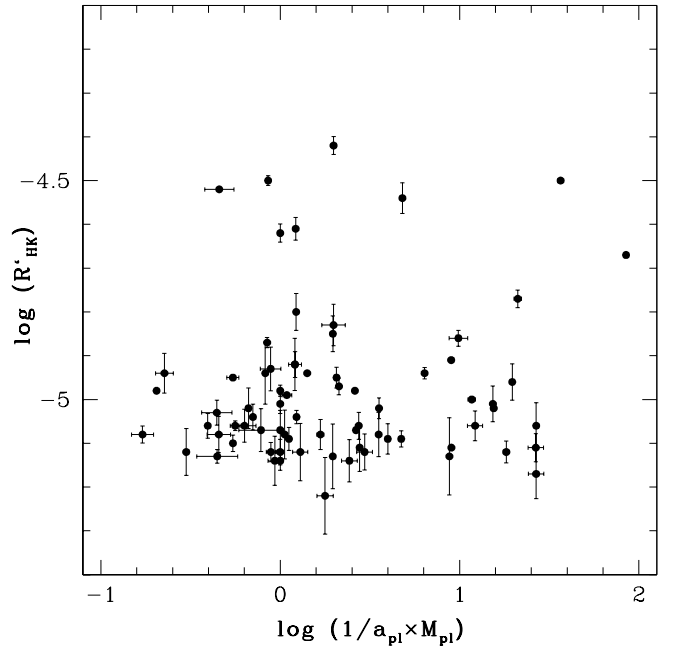}
\caption{Chromospheric activity indicator $\log (R'_{HK})$ as a function of the product of the planetary mass with the reciprocal semi-major axis $\log (1/a_{pl} \times M_{pl})$ for our sample of stars with planets.}
\label{default}
\end{figure}

Following the same strategy used by  Poppenhaeger et al. (2010) in their analysis of the coronal X-ray emission behavior in 
stars with planets, we show the distribution of the chromospheric activity indicator $\log (R'_{HK})$ of stars with planets as a function of planetary distance $\log (1/a_{pl})$ (Fig. 1). A close comparison of the distribution of $\log (R'_{HK})$ versus $\log (1/a_{pl})$ with the distribution of the coronal activity indicator $\log (\frac{L_X}{L_{bol}})$ versus $\log (1/a_{pl})$, illustrated in  Fig. 5 of Poppenhaeger et al. (2010), shows that, at least qualitatively, the behavior of the chromospheric activity indicator appears to be comparable to that of the coronal activity indicator.  The same aspect can be seen in Fig. 2 were we show the distribution of the chromospheric activity indicator $\log (R'_{HK})$ of stars with planets as a function of the product of the planetary mass with the reciprocal distance $\log (1/a_{pl})$.

\begin{figure}
\centering
\includegraphics[width=8.cm]{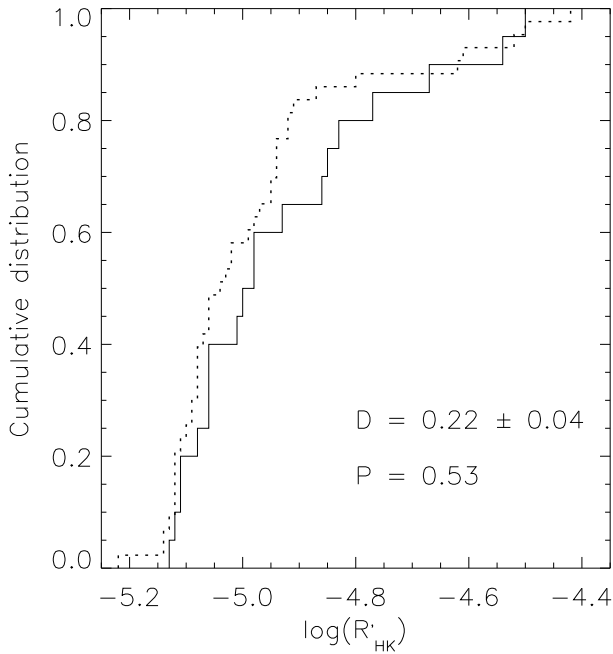}
\caption{The cumulative distributions of $\log (R'_{HK})$ for stars with $a_{pl} \le $ 0.2 AU (solid curve) and $a_{pl} \ge $ 0.5 AU (dashed curve).}
\label{default}
\end{figure}

\begin{table}[htb]									
\begin{center}									
\caption{Stellar parameters and chromospheric and coronal activity indicators for the sample of stars without planets (Schmitt 1997). }									
\begin{tabular}{lcccc} 									
\hline \hline									
Star	&	ST	&	$(B-V)$	&	$\log (R'_{HK})$	&	$\log (\frac{L_X}{L_{bol}})$	\\
\hline									
HD 4614		&	G0V...	&	0.587$\,\pm\,$0.003	&	-4.93$\,\pm\,$0.01	&	-6.59	\\
HD 10700		&	G8V		&	0.727$\,\pm\,$0.007	&	-5.01$\,\pm\,$0.02	&	-6.55	\\
HD 13974		&	G0V		&	0.607$\,\pm\,$0.005	&	-4.71$\,\pm\,$0.01	&	-5.13	\\
HD 14412		&	G8V		&	0.724$\,\pm\,$0.003	&	-4.89$\,\pm\,$0.04	&	-6.13	\\
HD 14802		&	G2V		&	0.608$\,\pm\,$0.017	&	-5.06$\,\pm\,$0.07	&	-5.06	\\
HD 16895		&	F7V		&	0.514$\,\pm\,$0.002	&	-4.92				&	-6.12	\\
HD 19373		&	G0V		&	0.595$\,\pm\,$0.007	&	-5.01$\,\pm\,$0.02	&	-6.83	\\
HD 30495		&	G3V		&	0.632$\,\pm\,$0.006	&	-4.61				&	-4.75	\\
HD 30652		&	F6V		&	0.484$\,\pm\,$0.003	&	-4.60				&	-5.03	\\
HD 69830		&	K0V		&	0.754$\,\pm\,$0.009	&	-4.98				&	-5.85	\\
HD 72673		&	K0V		&	0.780$\,\pm\,$0.010	&	-4.98$\,\pm\,$0.04	&	-5.84	\\
HD 90839		&	F8V		&	0.541$\,\pm\,$0.008	&	-4.83				&	-5.53	\\
HD 101501	&	G8Vvar	&	0.723$\,\pm\,$0.014	&	-4.57				&	-5.04	\\
HD 102870	&	F8V		&	0.518$\,\pm\,$0.015	&	-4.90				&	-5.85	\\
HD 103095	&	G8Vp	&	0.754$\,\pm\,$0.009	&	-4.88$\,\pm\,$0.02	&	-6.48	\\
HD 104304	&	K0IV		&	0.760$\,\pm\,$0.021	&	-4.95				&	-6.09	\\
HD 109358	&	G0V		&	0.588$\,\pm\,$0.009	&	-4.92$\,\pm\,$0.03	&	-6.44	\\
HD 114710	&	G0V		&	0.572$\,\pm\,$0.006	&	-4.75				&	-5.58	\\
HD 115617	&	G5V		&	0.709$\,\pm\,$0.007	&	-5.07				&	-6.53	\\
HD 117176	&	G5V		&	0.714$\,\pm\,$0.007	&	-5.02				&	-7.66	\\
HD 131156	&	G8V+...	&	0.720$\,\pm\,$0.015	&	-4.37				&	-4.62	\\
HD 141004	&	G0Vvar	&	0.604$\,\pm\,$0.008	&	-4.98$\,\pm\,$0.02	&	-6.05	\\
HD 142860	&	F6V		&	0.478$\,\pm\,$0.006	&	-4.76				&	-6.49	\\
HD 144579	&	G8V		&	0.734$\,\pm\,$0.001	&	-5.00$\,\pm\,$0.01	&	-6.19	\\
HD 161797	&	G5IV		&	0.750$\,\pm\,$0.015	&	-5.15$\,\pm\,$0.06	&	-6.32	\\
HD 185144	&	K0V		&	0.786$\,\pm\,$0.007	&	-4.88$\,\pm\,$0.06	&	-5.81	\\
\hline									
\end{tabular}									
\end{center}									
\end{table}									

In addition, still following the strategy proposed by Poppenhaeger et al. (2010), we carried out a statistical analysis of both $\log (R'_{HK})$ and $\log (\frac{L_X}{L_{bol}})$ activity indicators, of stars with close-in planets, namely planets with $a_{pl} \le $ 0.20 AU, searching for correlations with planetary mass and semi-major axis. Table 3 shows the result of such an analysis, with the Spearman's $\rho$ rank correlation for various combinations of stellar quantities and planetary parameters, for two sample: the first is composed of 19 stars with X-ray lumisonity from Kashyap et al. (2008), and the second one is composed of 13 stars with X-ray luminosities from Poppenhaeger et al. (2010). The first interesting aspect from this analysis is a possible anticorrelation between the chromospheric activity indicator and the planetary semi-major axis in stars hosting planets with $a_{pl} \le $ 0.20 AU for both samples. When we consider the entire sample of stars, we observe an absence of any significant correlation between these two parameters, 
without discrimination of $a_{pl}$ values.

\begin{table*}[ht]
\caption{Statistical analysis of chromospheric activity indicator $\log (R'_{HK})$ and planetary parameters and between coronal activity indicator $\log (\frac{L_X}{L_{bol}})$ for stars with planets with semi-major axis $a_{pl} \le $ 0.2 AU for two sample: one composed of 19 stars with X-ray lumisonity from Kashyap et al. (2008), and other composed of 13 stars with X-ray luminosities from Poppenhaeger et al. (2010).  }
\centering
\begin{tabular}{lcccc} 
\hline \hline
Parameters								&	Spearman's $\rho$	&	Probability {\it{p}}	&	Spearman's $\rho$	&	Probability {\it{p}}		  				\\
	          								&	\multicolumn{2}{c}{{\it (Kashyap et al. 2009)}}	&	\multicolumn{2}{c}{{\it (Poppenhaeger et al. 2010)}}					\\ \hline	
$\log R'_{HK}$ with $\log {\bf a_{pl}}$			&  -0.26$\pm$0.06 	&  0.30$\pm$0.12 		&  -0.29$\pm$0.06		&  0.34$\pm$0.10							\\
$\log R'_{HK}$ with $\log M_{pl}$				&  0.13$\pm$0.07 	&  0.61$\pm$0.19		&   0.20$\pm$0.08		&  0.52$\pm$0.16							\\
$\log R'_{HK}$ with $\log 1/a_{pl} \times M_{pl}$	&  0.17$\pm$0.06 	&  0.51$\pm$0.17		&   0.32$\pm$0.06		&  0.29$\pm$0.09							\\
$\log R'_{HK}$ with $\log \frac{L_X}{L_{bol}}$		&  0.64$\pm$0.09 	&  0.008$\pm$0.015	 	&   0.92$\pm$0.04		&  5.0$\times$10$^{-5}$$\pm$2$\times$10$^{-4}$	\\
\hline
\end{tabular}
\end{table*}

\begin{figure}
\centering
\includegraphics[width=9cm]{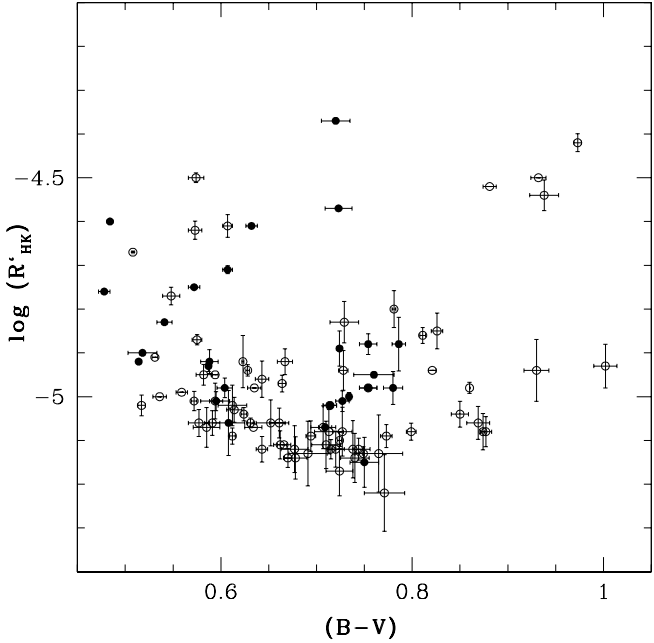}
\caption{Chromospheric activity indicator $\log (R'_{HK})$ as a function of colour index $(B-V)$. Open circles denote stars with planets and solid circles stars without planets.}
\label{default}
\end{figure}

\begin{figure}[hbt]
\centering
\includegraphics[width=8.cm]{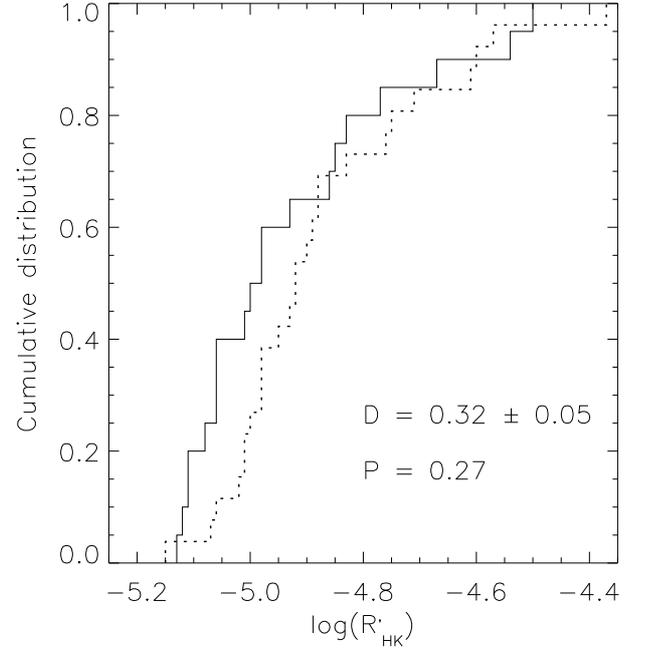}
\caption{The cumulative distributions of $\log (R'_{HK})$ for stars with planets (solid curve) and stars without planets (dashed curve). For the sample of stars hosting planets, we are only considering stars with planets with $a_{pl}\le0.20$ AU.}
\label{default}
\end{figure}

Two possible correlations are found in the present analysis: One of planetary mass $\log (M_{pl})$ with $\log (R'_{HK})$ and the other of product of the planetary mass with the reciprocal distance $\log (1/a_{pl} \times M_{pl})$ with $\log (R'_{HK})$. Stars with giant and close-in planets exhibit higher $\log (R'_{HK})$ values than those with small outermost planets, a result corroborating that obtained by Poppenhaeger et al. (2010) for the relationship between X-ray luminosities and planetary parameters. In this context, the Spearman coefficients from the correlation study of  $\log (R'_{HK})$ with  $\log (\frac{L_X}{L_{bol}})$, also displayed  in Table 3,  show a significant correlation between chromospheric and coronal activity indicators for stars with close-in planets.  This correlation indicates that the relations found by Poppenhaeger et al. (2010) between the coronal activity indicator and the planetary parameters must be the same obtained in our analysis of stellar chromospheric activity and the presence of planets around a star.

As an additional statistical test to verify if the present data sets, seggregated by planetary distances, are signiÞcantly different from one another, 
we applied the Kolmogorov-Smirnov test (Press et al. 1992), which calculates the probability that two distributions are derived 
from the same parent distribution. According to the K-S test, the {\it p} value indicates the probability that both distributions come from the same origin.
We conducted a K-S analysis, taking into account the $\log (R'_{HK})$ of stars with planets within $a_{pl} \le $ 0.20 AU (20 stars) and stars with planets 
beyond $a_{pl} \ge $ 0.50 AU (43 stars), once again following the strategy adopted by Poppenhaeger et al.  (2010).  Fig. 3 shows the cumulative functions for both $\log (R'_{HK})$ distributions. The errors were calculated considering the errors in $ \ log (R'_ {HK}) $ by applying random fluctuations with Gaussian distribution. The final values of {\it D} are the average of the standard deviation after the application of the fluctuations. The probability value of about  53\% obtained on the K-S test is consistent with the two distributions being drawn from the same population.  In agreement with  Poppenhaeger et al.  (2010), in their analysis of coronal X-ray emission,  the present result indicates that both distributions of chromospheric activity indicator  $\log (R'_{HK})$ of stars with close-in and far-out planets are very similar. 

\subsection{Comparison of chromospheric activity indicator in stars with and without planets} 

In a search for systematic differences, we compared the behavior of  chromospheric activity indicator $\log (R'_{HK})$ 
in stars with and without detected planets. In Fig. 4 we show $\log (R'_{HK})$ versus colour index ($B-V$) for both samples of stars, with no discernible differences. Both samples tend to follow the well known behavior of  $\log (R'_{HK})$, characterized by a decreasing of chromospheric activity with increasing color index.  In addition, in order to check for systematic differences between stars with and without detected planets, we compared the distributions of  $\log (R'_{HK})$ for the two aforementioned samples, considering only stars hosting close-in planets, in other words, stars with planets with $a_{pl} \le 0.20$ AU. Fig. 5 shows the cumulative functions for both distributions. A K-S test revels that the probability of both samples being drawn from the same parent distribution is 27\%, reinforcing the previous scenario, which shows no clear evidence of enhanced chromospheric activity associated to the presence of planetary companions. 

\section{Conclusions} 

We analyzed a sample of 74 stars with planets in the solar neighborhood, presenting chromospheric activity indicator $\log (R'_{HK})$, searching for possible effects of star-planet interaction on the stellar chromosphere. From these analysis, we found no signiÞcant correlations between the chromospheric activity indicator  $\log (R'_{HK})$ and planetary parameters: semi-major axis and product of the planetary mass with the reciprocal semi-major axis. However, we found a possible correlation between the $\log (R'_{HK})$ and mass and between the product of planetary 
mass and reciprocal semi-major axis, indicating that massive close-in planets are often found around stars with an enhanced chromospheric activity 
indicator. Such a result supports that obtained by Poppenhaeger et al. (2010) in their analysis of X-ray coronal luminosity in stars 
with planets. According to these authors, this dependence can be ascribed to selection effects, since the Doppler method for planet detection 
favors small and far--out planets around stars with low activity. Indeed, the present analysis shows a strong correlation 
between chromospheric and coronal activity indicators of stars with planets. Additionally, a statistical comparison between the $\log (R'_{HK})$ indicator in stars with 
and without detected planets shows no clear evidence of enhanced chromospheric activity associated to the presence of planetary companions. 
In summary, our analysis reveal no clear evidence of enhanced planet-induced activity in the chromosphere of the stars. In agreement with the conclusions drawn 
by Poppenhaeger et al. (2010) in their analysis of the X-ray luminosity behavior of stars with planets, any trends observed in the 
present study seem to be mostly the result of selection effects.  

\begin{acknowledgements}
Research activities at the Stellar Board of Universidade Federal do Rio Grande do Norte are supported by continuous grants from the
Brazilian agencies CNPq and FAPERN (J.R. De Medeiros and B. L. Canto Martins). M. L. Chagas and S. Alves acknowledge graduate fellowships from
the CAPES brazilian agency. L. P. de Souza Neto and I. C. Le\~ao acknowledge  fellowships of the CNPq brazilian agency.
We warmly thank the anonymous  referee for careful reading and suggestions that largely improved this paper. 
\end{acknowledgements}


\end{document}